\def\year{2021}\relax
\documentclass[letterpaper]{article} 
\usepackage{aaai21}  
\usepackage{times}  
\usepackage{helvet} 
\usepackage{courier}  

\usepackage{amsfonts}
\usepackage{bbm}
\usepackage{lineno}

\usepackage[hyphens]{url}  
\usepackage{graphicx} 
\usepackage{natbib}  
\usepackage{caption} 
\usepackage{amsmath}

\title{Vertical-Horizontal Structured Attention for Generating Music with Chords}
\author{Yizhou Zhao{\normalfont\textsuperscript{1}}, Liang Qiu{\normalfont\textsuperscript{1}}, Wensi Ai{\normalfont\textsuperscript{2}}, Feng Shi{\normalfont\textsuperscript{1}}, Song-Chun Zhu{\normalfont\textsuperscript{1}} \\
\textsuperscript{1}UCLA Center for Vision, Cognition, Learning, and Autonomy \\
\textsuperscript{2}University of California, Los Angeles \\
\texttt{\{yizhouzhao, liangqiu,va0817\}@ucla.edu, \{shi.feng, sczhu\}@cs.ucla.edu}}
\begin{document}
\maketitle

\begin{abstract}
In this paper, we propose a lightweight music-generating model based on variational autoencoder (VAE) with structured attention. Generating music is different from generating text because the melodies with chords give listeners distinguished polyphonic feelings. In a piece of music, a chord consisting of multiple notes comes from either the mixture of multiple instruments or the combination of multiple keys of a single instrument. We focus our study on the latter. Our model captures not only the temporal relations along time but the structure relations between keys. Experimental results show that our model has a better performance than baseline MusicVAE in capturing notes in a chord. Besides, our method accords with music theory since it maintains the configuration of the circle of fifths, distinguishes major and minor keys from interval vectors, and manifests meaningful structures between music phrases. 

\end{abstract}

\begin{figure*}[]
\label{fig:first}
\begin{center}
\centerline{\includegraphics[width=0.8\linewidth]{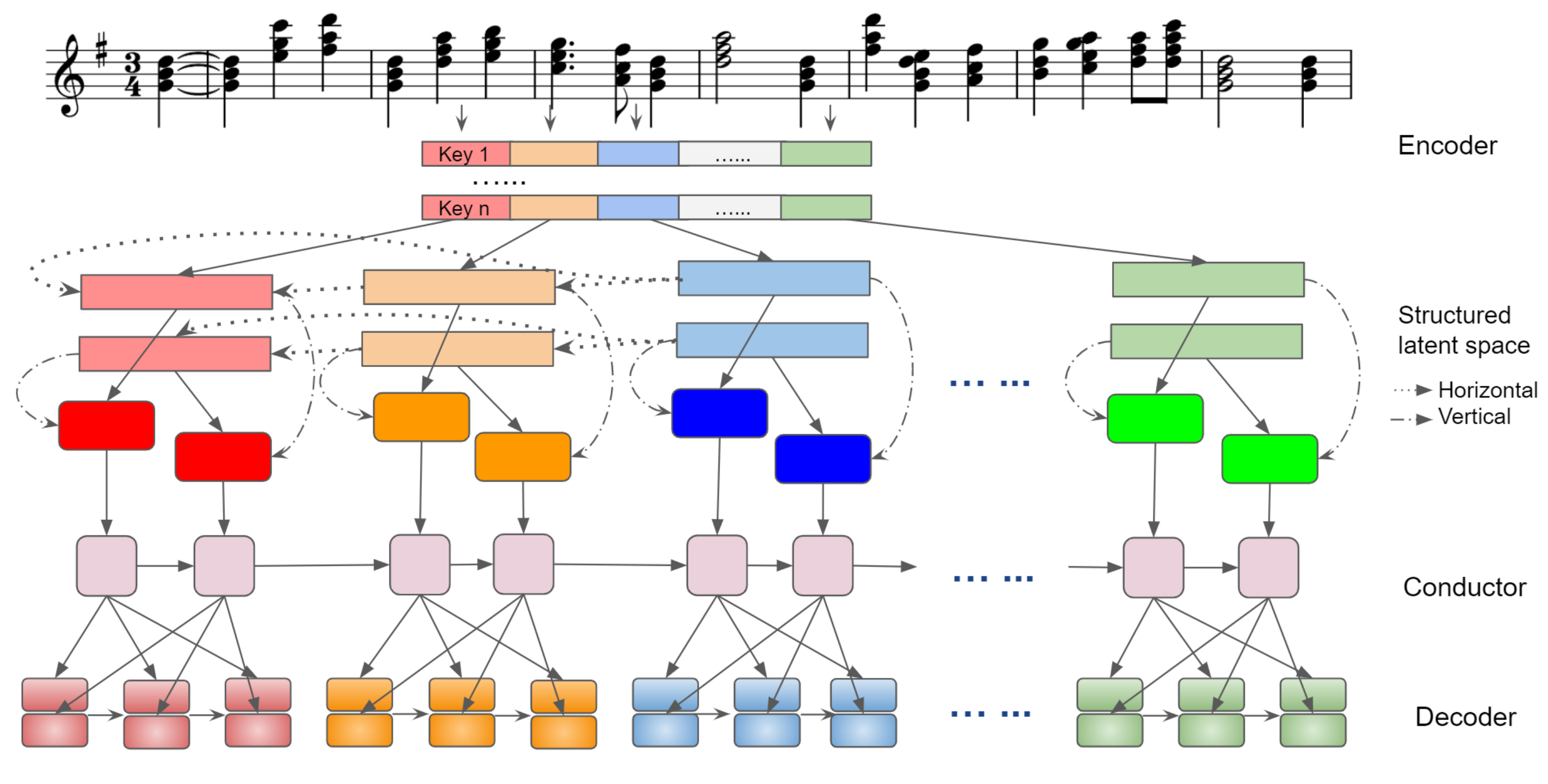}}
\caption{An overview of our Vertical-Horizontal VAE.}
\end{center}
\end{figure*}
\section{Introduction}
How does artificial intelligence (AI) inspire creativity? Recent years have witnessed the rise of AI art in literature \cite{roemmele2016writing}, paintings \cite{davis2016co} and music \cite{chuan2018modeling}. The temporal quality of music makes it different from others: the art of sound expresses ideas and emotions through the elements of rhythm and melody. A large number of deep neural network models for music generation have been proposed over the years \cite{huang2018music,donahue2019lakhnes,dong2018musegan}. Some researchers apply models such as convolutional neural networks (CNNs) \cite{DBLP:conf/ismir/YangCY17} and Transformers \cite{huang2018music} to generate music. Our work employs recurrent neural networks (RNNs) to do the task because of their strength in handling sequential data such as speeches \cite{qiu2018non} and dialogues \cite{10.5555/3298023.3298047}.

To compose coherent musical pieces, musicians control rhythms and melodies by repeating, shifting and varying music notes to bring enjoyment, surprise, and beauty. Researchers have realized that the attention mechanism helps capture the referential phenomenon between bars (music bar lines) and audio tracks. For example, Music Transformer shows that self-attention is well-suited for modeling music \cite{huang2018music}. This paper presents a lighter and easier solution to capture the temporal and spatial relation between notes. We introduce vertical-horizontal VAE (VH-VAE), a model based on the structure attention network (SAN) \cite{kim2017structured} and MusicVAE \cite{conf/icml/RobertsERHE18}. SAN extends the basic attention procedure on networks, which can capture structural dependencies without abandoning end-to-end training, and MusicVAE is a hierarchical extension of variational RNN (VRNN) \cite{chung2015recurrent} on its application to music. As an art of time, music satisfies certain criteria of coherence related to rhythm, tension, and emotion flow. We model it by horizontal structured attention that formulates linkage between bars. As an art with harmonic characteristics, music notes play simultaneously as a chord to add texture to a melody. We model it by vertical structured attention that passes messages between keys. Besides, the chord makes the music generation a multi-label problem. We propose Permutation Loss base on Focal Loss \cite{lin2017focal} to overcome the limitation of a single selection of one note per time. 

As deep learning is receiving growing attention as an approach for music generation, several issues including creativity, interactivity, and theory awareness in music information retrieval are proposed by recent research groups.  We may be not satisfied if the generated musical content tends to mimic the training set without exhibiting true creativity \cite{briot2020deep}, and we expect that human users can compose cooperatively with a machine \cite{donahue2019piano}. However, how to evaluate creativity is mathematically hard to define and it often needs user's studies to judge whether a machine cooperates properly in composing music. Recently, music theory has obtained attention in deep learning models \cite{jaques2017tuning, brunner2017jambot}. The theory tells that groups of notes belong to keys, chords follow progressions, and songs have consistent structures made up of musical phrases \cite{jaques2017tuning}. We propose to conduct extensive experiments on theory-aware analysis for music-generating models, in order to test whether they can show some aesthetic patterns from a music theory perspective.

We train our model on the MAESTRO dataset \cite{hawthorne2018enabling}. Results show that our model performs better than the state-of-the-art models as our model captures at least $10\%$ more chords. Learning from masterpieces, our model draws a similar picture of the circle of fifths and distinguishes the patterns between minor and major chords from interval vectors. Besides, the visualization of the tree-structured attention can be a guide for teaching and composing music. 

\section{Related work}
A large number of deep neural network models have been proposed for music generation from audio waves \cite{DBLP:conf/iclr/MehriKGKJSCB17,hawthorne2018enabling, engel2017neural}, or note sequences \cite{brunner2017jambot, DBLP:conf/ismir/BrunnerKWW18}. MelodyRNN \cite{DBLP:journals/tcyb/WuHWHZ20} is a straightforward RNN symbolic-domain music generator with the variants that aim to learn longer-term structures in musical pieces. MIDINet \cite{DBLP:conf/ismir/YangCY17} is the CNN version with a generator and discriminator, making it a generative adversarial network (GAN). MuseGAN \cite{dong2018musegan} is another model based on CNN and GAN for multi-track music. Recently, Music Transformer \cite{huang2018music} and its variants such as LakhNES \cite{donahue2019lakhnes} are proposed to generate minute-long compositions with help from self-attention.

VAE \cite{kingma2013auto} and its recurrent version VRNN \cite{chung2015recurrent} are viewed as an important contribution to fill the gap between neural networks and traditional Bayes.
MusicVAE \cite{conf/icml/RobertsERHE18} is a hierarchical VAE that learns a summarized representation of musical qualities as a latent space. MIDIMe \cite{dinculescu2019midime} improves it by resembling the structure of the input melody. MIDI-VAE \cite{DBLP:conf/ismir/BrunnerKWW18} brings more inputs channels including pitches, instruments, and velocities, making it efficient in style transfer. JamBot \cite{brunner2017jambot} bridges the gap between deep neural networks and music theory related to chords by predicting chords progression from chords embedding.

SANs \cite{kim2017structured} were proposed as a generalization of the basic attention procedure \cite{vaswani2017attention}. They allow attention mechanisms beyond the standard soft-selection approach to incorporate certain structured inductive biases. Especially, hierarchical structures like constituency trees \cite{ijcai2020-560} have been shown to be effective for various natural language processing tasks.

Our idea starts from MusicVAE, but abandons its heavy networks and incorporates a tree-structured attention network. Compared with regular attention mechanism \cite{luong-etal-2015-effective}, the tree-structured attention is much faster to calculated \cite{kim2017structured}, resulting in a lightweight model easier to train, even on a single CPU. Most importantly, it depicts tree-shaped dependencies along bars and between keys, which is more intuitive for musical analysis and composition.

In summary, this paper makes the following con-tributions.  (1) We propose to incorporate a structured  attention mechanism within MusicVAE to capture the relationships between bars and keys. (2) We introduce Permutation Loss to generate chords. (3) We conduct extensive experiments to show that our model is theory-aware.
\section{Problem Formulation}
Our main task is to reconstruct music files in the MIDI format (a symbolic representation of music that resembles sheet music) from a latent variable model. MIDI files contain multiple tracks with multiple instruments. We focus our formulation on one track in the beginning and expect to extend it to multiple tracks in future applications. 

For simplicity, we fix the length of music pieces and assume all of them have the same number of bars with length $N_T$. We consider the smallest note as the sixteenth and ignore the velocity of notes.

We factorize the problem of generating music notes from latent space to two parts: an encoder for labeling the note sequence history with a sequence of latent states, and a decoder to generate the next pitches from the latent state. Suppose $N_P$ is the size of pitch space $P$, a standard single track MIDI music sheet $X$ is a sequence of note pitches $(x_1, x_2,..., x_T)$, where $x_t \in \{0, 1\}^{N_P}$ indicating the pitches played at time $t$.  Then each $x_t$  belongs to a latent state $z_t$, summarizing the effects of previous notes and indicating the pitches corresponding to the next step. In the VRNN framework, each time step contains a VAE. For sequential data, the parameterization of the generative model is factorized by the posterior $ p\left(z_{t} | x_{<t}, z_{<t}\right)$ and the generative model $p\left(x_{t} | z_{\leq t}, x_{<t}\right)$, i.e.
\begin{equation}\label{eq:1}
    \begin{aligned}
        p(x \leq T, z \leq T)=\prod_{t=1}^{T} &\bigg[ p\left(x_{t} | z_{\leq t}, x_{<t}\right)p\left(z_{t} | x_{<t}, z_{<t}\right)\bigg]
    \end{aligned}
\end{equation}
In a MIDI file with multiple tracks (or instruments), the note $x_t \in \{0, 1\}^{N_P\times N_I}$ can be decomposed as $(x_{t,1}\oplus x_{t,2}\oplus ... \oplus x_{t,N_I})$ where $N_I$ is number of instruments or audio tracks. The direct sum $\oplus$ suggests that the multi-track scenario can be handled well if the tracks are trained independently and then merged together \cite{dong2018musegan}.  Another harder scenario we consider in this paper is a single track with chords. In such case, $x_t \in \{0,1\}^{N_P \times N_K}$ can be decomposed into $(x_{t,1} + x_{t,2} + ... + x_{t,N_K})$, where $N_K$ is the maximum number of the pitches in a chord (or keys pressed on a musical instrument) at one time. Because of the large number of the combination of keys to produce a chord and the uncertain number of keys pressed at each time, researchers often bypass this problem by turning the chords into arpeggios or just considering the most frequent chords \cite{DBLP:conf/ismir/BrunnerKWW18}. Hence, We desire to solve the aforementioned problem with a direct method and generalize composing chords as the multi-label classification problem. Meanwhile, we apply the structured attention mechanism to illustrate the self-referential essence of music. 

\begin{figure}[h]
  \centering
  \includegraphics[width=\linewidth]{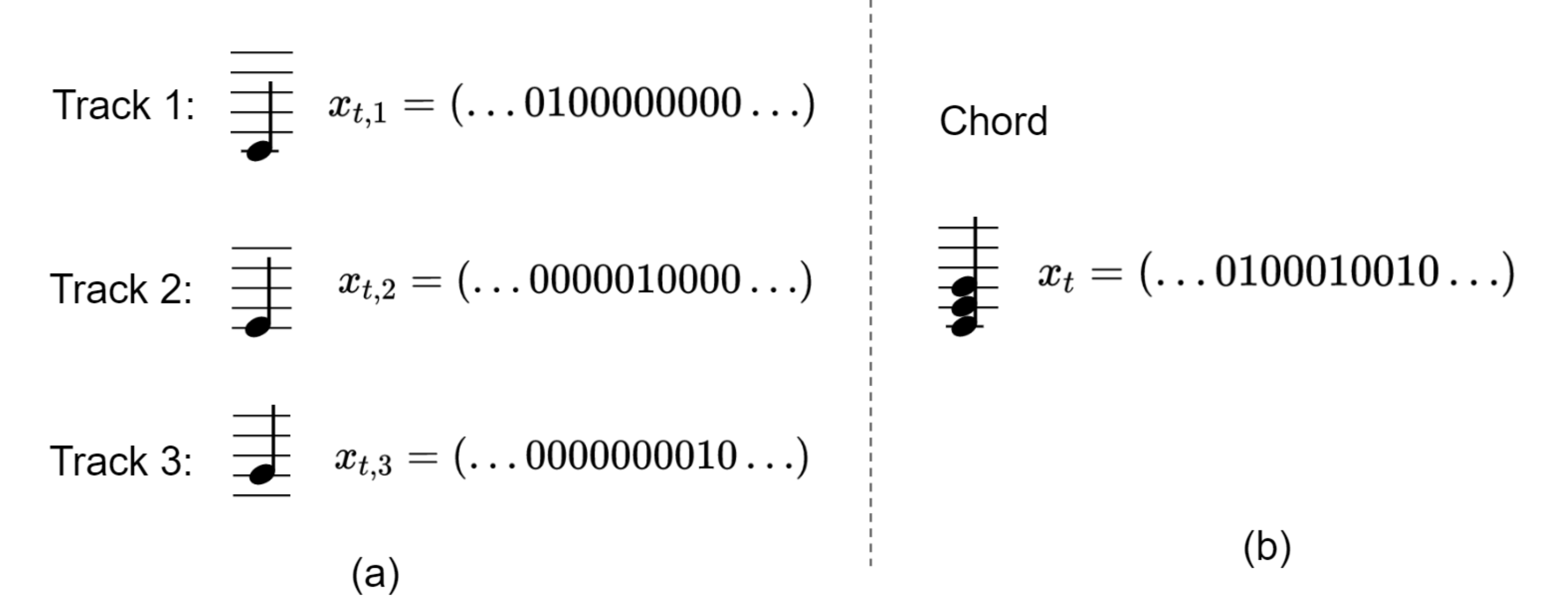}
  \caption{A comparison between notes in multiple tracks and a chord in a single track: (a) notes in multiple tracks; (b) a chord in one track}
  \label{fig:chord_track}
\end{figure}

The comparison between the above two scenarios is shown in figure \ref{fig:chord_track}. A regular MIDI file is often the compound of those two: multiple instruments generate chords. Our study focuses on the study of the chords in a single track, which can be applied to generate a solo. 
\section{Preliminaries}
\subsection{Variational Recurrent Neural Network}
RNNs are able to represent dependencies in sequential data, by adapting and propagating a hidden state. The introduced state space models with delineating layers are beneficial for making efficient posterior inferences. Derived from VAEs, VRNNs merge the generative model with RNN, which makes them possess the ability to generate sequential data. 


In a VAE, the latent code $z$ is a random vector sampled from a prior $p(z)$ and the data generation model is described by $p(x|z)$. The VAE consists of an encoder $q_\lambda(z|x)$ for approximating the posterior $p(z|x)$, and a decoder $p_\theta(x|z)$ for representing the distribution $p(x|z)$. The principle of the variational inference maximizes the likelihood by maximizing evidence lower bound (ELBO):
\begin{equation}\label{eq:2}
\mathbb{E}\left[\log p_{\theta}(x | z)\right]-\text{KL}\left(q_{\lambda}(z | x) \| p(z)\right) \leq \log p(x)
\end{equation}
where $\text{KL}$ stands for the Kullback–Leibler divergence.

The VRNN contains a VAE at each time step. For sequential data, the parameterization of the generative model is factorized by the posterior $ p\left(z_{t} | x_{<t}, z_{<t}\right)$ and the generative model $p\left(x_{t} | z_{\leq t}, x_{<t}\right)$. Equation \ref{eq:1} shows the likelihood function for the entire model. The learning objective function becomes maximizing the ELBO for all time steps:
\begin{equation}
\begin{split}
\text{ELBO} =\mathbb{E} \Big\{ \sum_{t=1}^{T}(&\text{-}\mathrm{KL}(q(z_{t} | x_{\leq t}, z_{<t}) \| p(z_{t} | x_{<t}, z_{<t}))\\
&+\log p(x_{t} | z_{\leq t}, x_{<t})) \Big\}
\end{split}
\label{eq:elbo_time_steps}
\end{equation}
In music generation, $x_t$ represents the pitches at time $t$, and the gradient of $z_t$, which usually has a Gaussian prior, is sampled by the reparameterization trick \cite{kingma2013auto}.

\subsection{Non-projective Dependency Tree Attention}
Our model employs a non-projective dependency tree attention layer with VRNN to learn the structure in notes generation scenario. In such a layer, we use self-attention over the sentence embeddings from the encoder so that no explicit query from the decoder is required. The potentials $\theta_{i,j}$, which reflect the score of selecting the $i$-th sample $y_i$ being the parent of the $j$-th sample $y_j$ in a data sequence $y=\{y_1,\ldots,y_n\}$ with length $n$, are calculated by
\begin{equation}\label{eq:4}
    \theta_{i,j}=\tanh\left(s^T \tanh(W_1 h_i + W_2 h_j + b)\right)
\end{equation}
where $s,b,W_1,W_2$ are parameters and $h_i, h_j$ are the encodings of $y_i, y_j$.    
The probability of a parse tree $r$ is
\begin{equation}\label{eq:5}
            p(r|y)=\textrm{softmax}\left(\sum_{i\neq j}\mathbbm{1}\{r_{i,j}=1\}\theta_{i,j}\right)
\end{equation}
where the latent variable $r_{i,j}\in\{0,1\}$ for all $i\neq j$ indicates that the $i$-th sample is the parent of the $j$-th sample. 
It is possible to calculate the marginal probability of each edge $p(r_{i,j}=1|y)$ for all $i,j$ in $O(n^3)$ time using the inside-outside algorithm \cite{covington2001fundamental}.

Then the soft-parent or the context vector of the $j$-th sample is calculated using parsing marginals, i.e., 
\begin{equation}
    c_j=\sum_{i=1}^n p(r_{i,j}=1|y) y_i
\end{equation}
The original embedding is concatenated with its context vector to form the new representation 
\begin{equation}
    \hat{y}_j=[y_j;c_j]
\end{equation}
The new representation $\{\hat{y}_1,\ldots,\hat{y}_n\}$ are attended over using the standard attention mechanism at each decoding step by an LSTM decoder. 

\section{Vertical-Horizontal VAE}
Figure 1 provides a graphical illustration of our model processing a particular musical sequence. The input sequence is first fed to the LSTM encoder to generate the latent vectors, after which structured attention is applied to both keys (vertically) and to bars (horizontally) on the latent vectors to generate contexts for each measure and key. These contexts are combined with the previous latent variables then autoregressively passed through a conductor to produce the initial input of the LSTM decoder, which finally generates the output sequence.

\subsection{Vertical-horizontal Attention}
The hierarchical RNN for the decoder was proposed by MusicVAE, which assumes the input sequence $X$ can be segmented into $N_U$ nonoverlapping subsequences: \begin{align}
X &= \{y_{1}, y_{2}, \ldots, y_{N_U}\} \\ 
    \text{and~~} y_{u} &= \{ x_{t_{u}}, x_{t_{u}+1}, x_{t_{u}+2}, \ldots, x_{t_{u+1}-1} \}
\end{align} 
Usually, each $y_u$ represents the notes in the $u$-th bar; $t_{u}$ stands for its starting time tick, and $t_{u+1}-1$ the ending. In the multi-key situation, we further assume that
\begin{align}
    y_u &= y_{u,1} + y_{u,2} + ... + y_{u,N_K} \\
    \text{and~~} y_{u,j}&= x_{t_u,j} + x_{(t_u + 1),j} + ... + x_{(t_{u+1}-1),j}
\end{align}
Recall that $N_K$ is the number of keys and $x_{i,j}$ stands for the pitch pressed for the $j$-th key at time $i$. Then $y_{u,j}$ is the pitch sequence produced by the $j$-th key in the $u$-th bar. The \textbf{vertical attention} is formulated as a tree attention across keys:
\begin{align}
    \hat{y}_{u,j}^v &= [y_{u,j}; \,c^v_{u,j}] \\
    c_{u,j}^v &= \sum_{i = 1}^{N_K} p(r^v_{i,j} = 1|y_{u})\cdot \,y_{u,i}
\end{align}
And label variable $r^v_{i,j}$ and potential $\theta^v_{i,j}$ are calculated from $y_{u,i}$ and $y_{u,j}$ according to Equation \ref{eq:4} and \ref{eq:5}. Similarly, the \textbf{horizontal attention} is formulated as a tree attention across bars:
\begin{align}
    \hat{y}^h_{u} &= [y_u; c^h_u] \\
    c_{u}^h &= \sum_{i = 1}^{u} p(r^h_{i,u} = 1|X)y_{i}
\end{align}
The label variable $r^h_{i,j}$ and potential $\theta^h_{i,j}$ are calculated from $y_{i}$ and $y_{j}$.

The combination of vertical and horizontal attention gives the new representation $\hat{y}_{u,j} = [y_{u,j}; c^v_{u,j}; c^h_u]$. Once we get this representation, it is passed through an LSTM as a conductor and then a decoder RNN to get the output notes. 

The encoder and decoder with a conductor share the structure as those in MusicVAE except that we apply a single-layer LSTM on each.

\subsection{Permutation Loss}
Like the serious imbalance problem in object detection between the number of labeled object instances
and the number of background examples \cite{liu2020deep}, the pitches distribution in music is also imbalanced: most of the pitches are rarely played, and a song usually contains one or several main melodies. Focal Loss (FL) \cite{lin2017focal} was proposed to address this problem by rectifying the cross-entropy loss, such that it down-weights the loss assigned to correctly classified examples:
\begin{equation}\mathrm{FL}\left(p_{\mathrm{t}}\right)=-\alpha_{\mathrm{t}}\left(1-p_{\mathrm{t}}\right)^{\gamma} \log \left(p_{\mathrm{t}}\right)\end{equation}
\begin{equation}p_{\mathrm{t}}=\left\{\begin{array}{ll}
p & \text { if } y=1 \\
1-p & \text { otherwise }
\end{array}\right.\end{equation}
where $\gamma$ and $\alpha$ are adjustable hyper-parameters. Its goal is to weaken the weight of the samples that the model has been able to predict well, and make the model concentrate on the hard cases. When $\gamma = 0, \alpha_t = 1$, Focal Loss is equivalent to the cross-entropy loss. i.e. the loss from generative model in ELBO.

In addition, for chord decomposition, to overcome the problem of producing overlapping pitches, we add \textbf{Permutation Loss} (PL) as penalization:
\begin{equation}
    \text{PL} = \sum_{k=2}^{N_K}\mathbb{E}_{\hat{p}_{1:k-1}}[\log p_k(x|z)]
\end{equation}
where $\hat{p}_{1:k-1}$ is the distribution of pitches produced by the top $k-1$ keys and $p_k$ is the pitch distribution for the $k$-th key. To prevent that one key highly prefers a certain pitch, we permute the order of keys randomly in the training steps. Notice that our target is to minimize PL so that it minimize the likelihood of the $k$-th key producing overlapping pitches produced by the previous $k-1$ keys. 

\section{Experiments}
We train our model on the MAESTRO (MIDI and Audio Edited for Synchronous TRacks and Organization) dataset \cite{hawthorne2018enabling}, which contains over a thousand-hour paired audio and MIDI recordings from nine years of International Piano-e-Competition events. We randomly select $90\%$ of the data for training and $10\%$ for testing in reconstruction task. For data prepossessing, we extract only the \textit{grand piano} track of the MIDI file and separate each track into pieces of the same length. 

Since the taste of music is subjective, it is difficult to evaluate the performance of the music generation model from a single perspective. We propose a method to evaluate the quality of the generative models for music. First, we must mathematically \textbf{make a comparison} between the quality of generated music pieces with original ones. Second, we can test whether musical pieces generated by the model cope with some \textbf{music theory}. Third, we may conduct \textbf{user experiments} or consult musicians to assess the quality of the generated music. Besides, \textbf{accessibility} which may include model size and training difficulty is considered.

Since our model is a derivative model of VAE, the quality of reconstructed notes is taken into account. We also compare the sizes between different models. Then, we devote our study to testing whether our model correlates with theories even though no constraints with respect to music theory are applied during the training steps. We have not conducted experiments on user experience due to the difficulty of selecting the experimental group and the control group. We leave it as future work.

\subsection{Reconstruction Quality}
Since we have ignored the velocity of notes, predicting the activeness of a certain pitch is a binary classification problem We borrow the idea of statistical hypothesis testing to form metrics to evaluate our model.
\begin{figure}[h]
  \centering
  \includegraphics[width=0.8\linewidth]{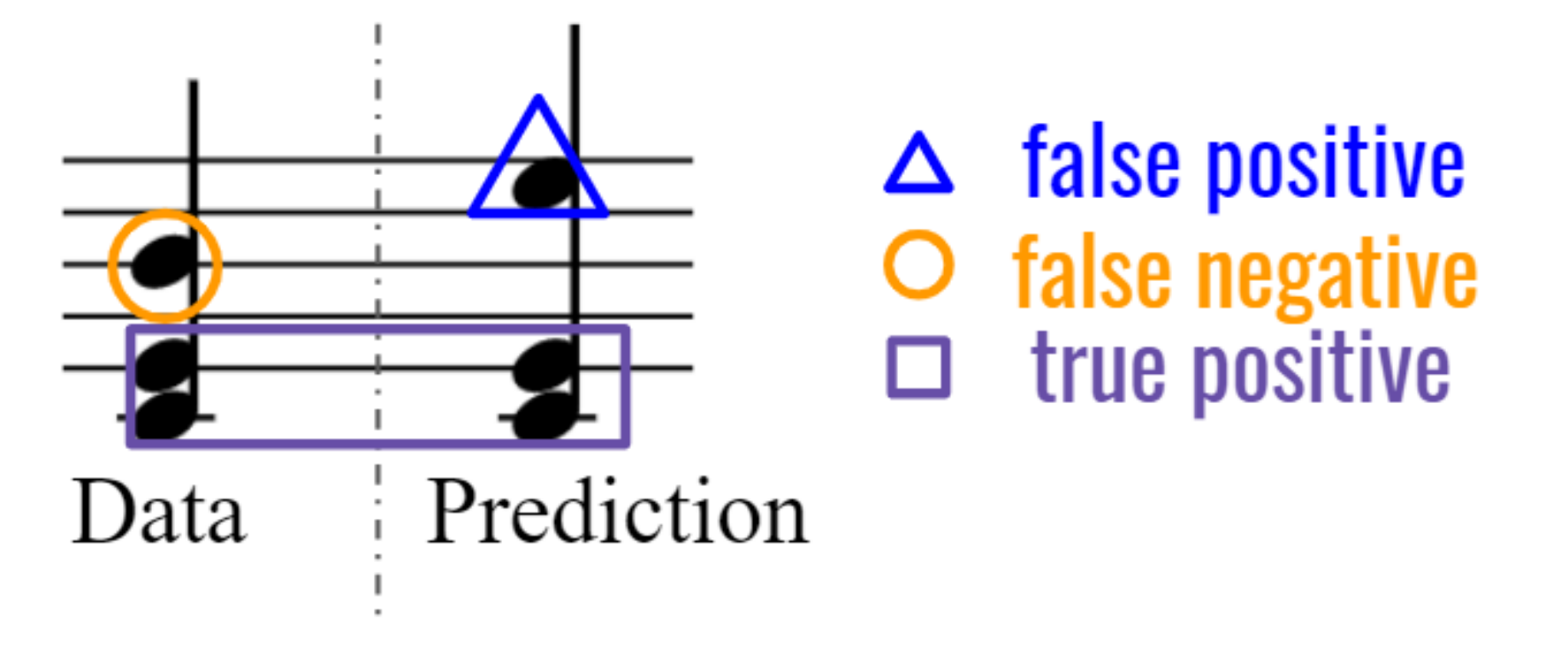}
  \caption{Statistical measures for chord reconstruction}
  \label{fig:2}
\end{figure}

Figure \ref{fig:2} illustrates the statistical measures to evaluate the quality of chord reconstruction. True positive rate (TPR) measures the proportion of actually activated pitches that are correctly identified as such, and positive predictive values (PPV) are the proportions of activated pitches that are true positive in the original pieces\footnote{Since most of the pitches are silent, negative predictive value (NPV) and true negative rate (TNR) are not considered as important measures.}.
\begin{table}[h]
\footnotesize
\centering
\begin{tabular}{ll|ll}
\hline
\textbf{Model}                            & \textbf{Bars} & \textbf{PPV}     & \textbf{TPR}     \\ \hline
MusicVAE (Single Key)             & 16         & 56.01\% & 14.14\% \\
Baseline (Two Keys)   & 16         & 51.40\% & 24.58\% \\
VH-VAE (Two Keys)                 & 16         & 53.03\% & \textbf{28.43\%} \\
Baseline (Three Keys) & 16         & 42.27\% & 23.33\% \\
VH-VAE (Three Keys)               & 16         & 47.44\% & 25.76\% \\ \hline
MusicVAE (Single Key)             & 32         & 52.89\% & 13.30\% \\
Baseline (Two Keys)   & 32         & 50.62\% & 27.20\% \\
VH-VAE (Two Keys)                 & 32         & 47.27\% & \textbf{31.79\%} \\
Baseline (Three Keys) & 32         & 38.33\% & 27.57\% \\
VH-VAE (Three Keys)               & 32         & 39.31\% & 28.62\% \\ \hline
\end{tabular}
\newline
  \caption{Model comparison by PPV and TPR}
  \label{tab:freq}
\end{table}

Masterpieces often contain a lot of chords. To reproduce the chords, it needs two steps: determine the number of keys and assign each key to the right pitch. Traditional \textit{Music + VAE} methods either broke chords into arpeggio or reduced the number of chords into a few. Those methods bring down the quality of the original pieces.
The original MusicVAE does not aim to reproduce chords. It is conservative to capture at least one pitch in the chord since it produces a single pitch at every time tick. Therefore, it has a high PPV (above $50\%$). However a low TPR (below 20\%) indicates it only captures a low percentage of the note pitches. 

We made modification to MusicVAE based on Tied-Parallel LSTM \cite{johnson2017generating} (a model to generate polyphonic music) to form our baseline. Specifically, we kept the model structure of MusicVAE but changed the loss function into binary cross-entropy loss.
 We also added the regular attention mechanism and Permutation Loss for the baseline to make it in line with our VH-VAE. Those \textit{Attention + MusicVAE} models generally have a lower PPV but a dramatic improvement on the TPR compared with MusicVAE. Adding our vertical-horizontal mechanism generally improves the PPV and TPR except for the model \textit{Attention + MusicVAE(Two keys)} with $32$ bars. Notice that even though our model spends $O(n^3)$ time calculating structured attention, applying the regular attention takes much more time since it requires calculating scores among all keys in all bars. When the number of bars increases from $16$ to $32$, Table \ref{tab:freq} shows that all models lose some performance in PPV; however the models with attention mechanism make a little improvement on TPR.


\subsection{Model Size Comparison}
As shown in Table \ref{tab:size}, we keep our model size small, making it easy to train. For \textit{JamBot} and \textit{Music Transformer}, even though there are no officially released models so far, we estimate the pretrained model sizes both as a few hundred megabytes from researchers who re-implemented their algorithms. Compared with MusicVAE, we shrink the input and output sizes of the layers of the encoder, conductor, and decoder, while the introduced structured attention mechanism only brings two sets of parameters $(s, b, W_1, W_2)$ for vertical and horizontal attention calculation.
\begin{table}[]
\begin{tabular}{l|l}
\hline
\textbf{Model} & \textbf{Model size} \\ \hline
JamBot \cite{brunner2017jambot}         & N/A                   \\
MidiNet \cite{DBLP:conf/ismir/YangCY17}           & $<10$ mb               \\
MIDI-VAE \cite{DBLP:conf/ismir/BrunnerKWW18}          & $<100$ mb              \\
MusicVAE \cite{conf/icml/RobertsERHE18}          & $>100$ mb              \\
MuseGAN \cite{dong2018musegan}           & $<100$ mb              \\
Music Transformer \cite{huang2018music} & N/A                   \\
VH-VAE         & $<10$ mb               \\ \hline
\end{tabular}
\caption{Model size comparison}
\label{tab:size}
\end{table}

\subsection{Theory-aware Analysis}
In this part, we analyze our model and test whether it maintains some theoretical features learned from masterpieces. 

\subsubsection{Circle of Fifths}
In Western music theory, an octave is usually divided into twelve notes under the twelve-tone equal temperament system. The circle of fifths provides a geometric interpretation of the relationships between these notes. Scales close to each other share similar keys with each other, and chords progressions are close to each other on the circle of fifths.
\begin{figure}[h]
  \centering
  \includegraphics[width=\linewidth]{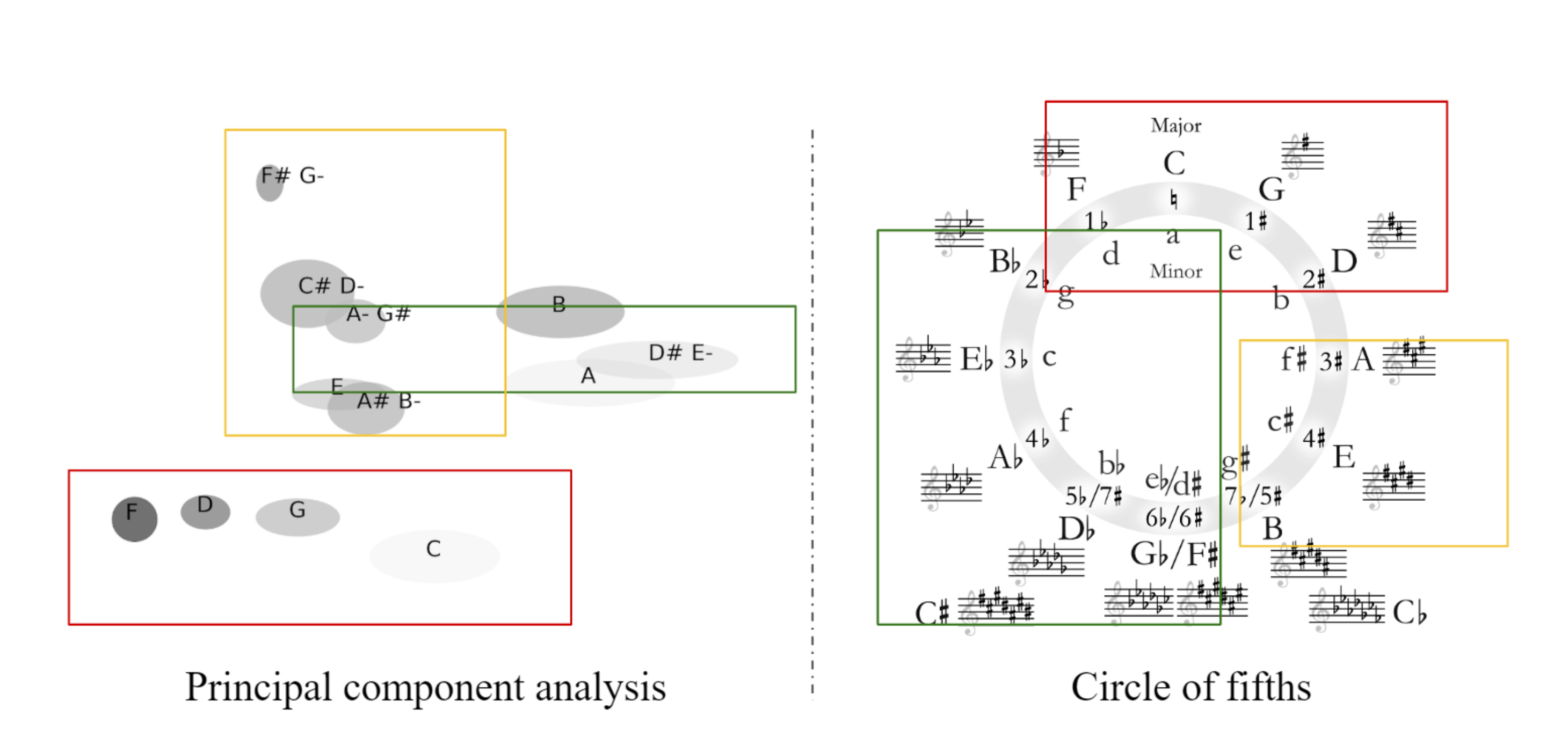}
  \caption{Circle of Fifths: machine and theory}
  \label{fig:circle}
\end{figure}

The result of our experiment is illustrated in Figure \ref{fig:circle}. By using the Principal Component
Analysis (PCA) on the latent vector $z$, we reduced our embedding of chords to two dimensions. Results show the forms of clustering, in which notes within the same cluster are generally close to each other on the circle of fifths. Considering that our model puts no constraints regarding music theory, and there is no explicit encoding for the model to extract features from chords, the result shows that our model can focus its attention on chords, and learn concepts of music theory regarding chord similarities. However, it seems that our deep learning model is somehow confused with $E$ and $E_\flat$, which triggers our interest in going deeper into the composition procedure for those two keys for further study.

\subsubsection{Interval vector}
\begin{figure}[h!]
  \centering
  \includegraphics[width=\linewidth]{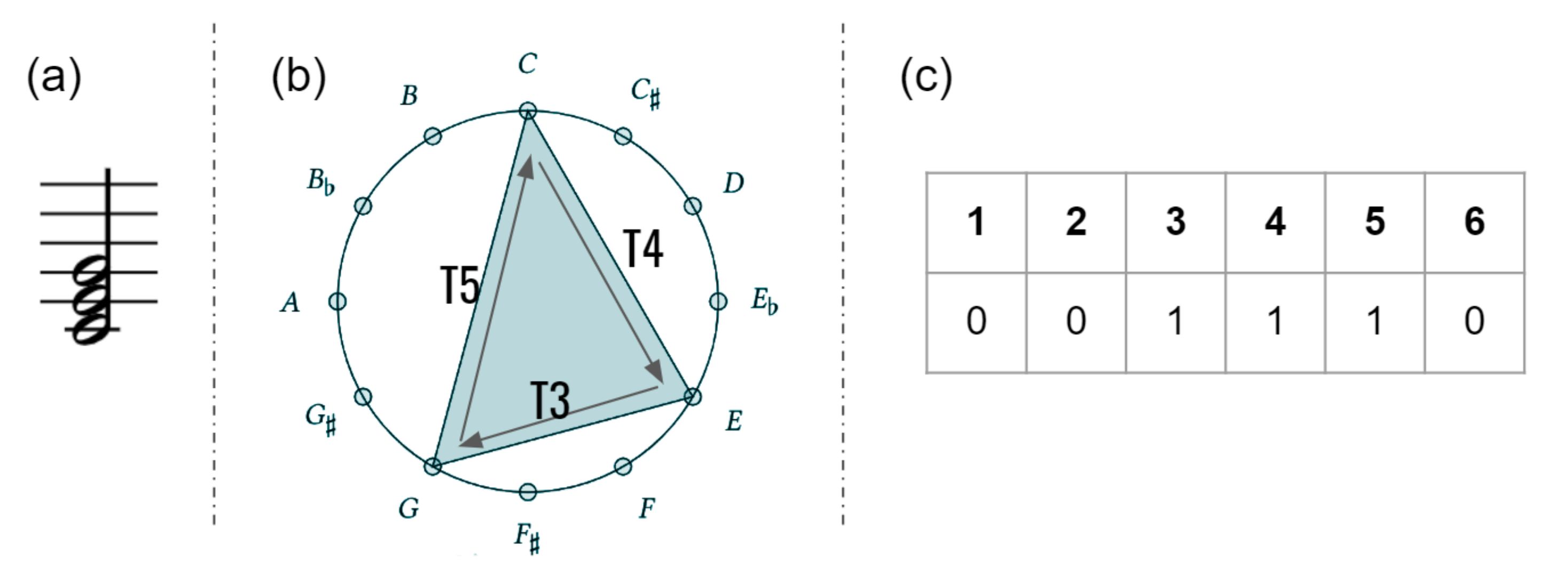}
  \caption{(a) Note of C major triad; (b) Intervals on $\mathbb{Z}_{12}$; (c) Interval vector: $(0,0,1,1,1,0)$}
  \label{fig:interval_vec}
\end{figure}

An interval vector, in music set theory, is an array of natural numbers that summarize the intervals present in a set of pitch classes. More precisely, for a given pitch-class set $S$, which has a bijection with $\mathbb{Z}_{12}$, the interval vector $IV$ of $S$ is an array of six coefficients $\{v_i\}_{i = 1,2,...,6}$, each one describing the number of times an interval of $i$ semitones appears in $S$. Figure \ref{fig:interval_vec} shows an example of how to calculate the interval vector of $C$ major triad.

\begin{figure}[h]
  \centering
  \includegraphics[width=\linewidth]{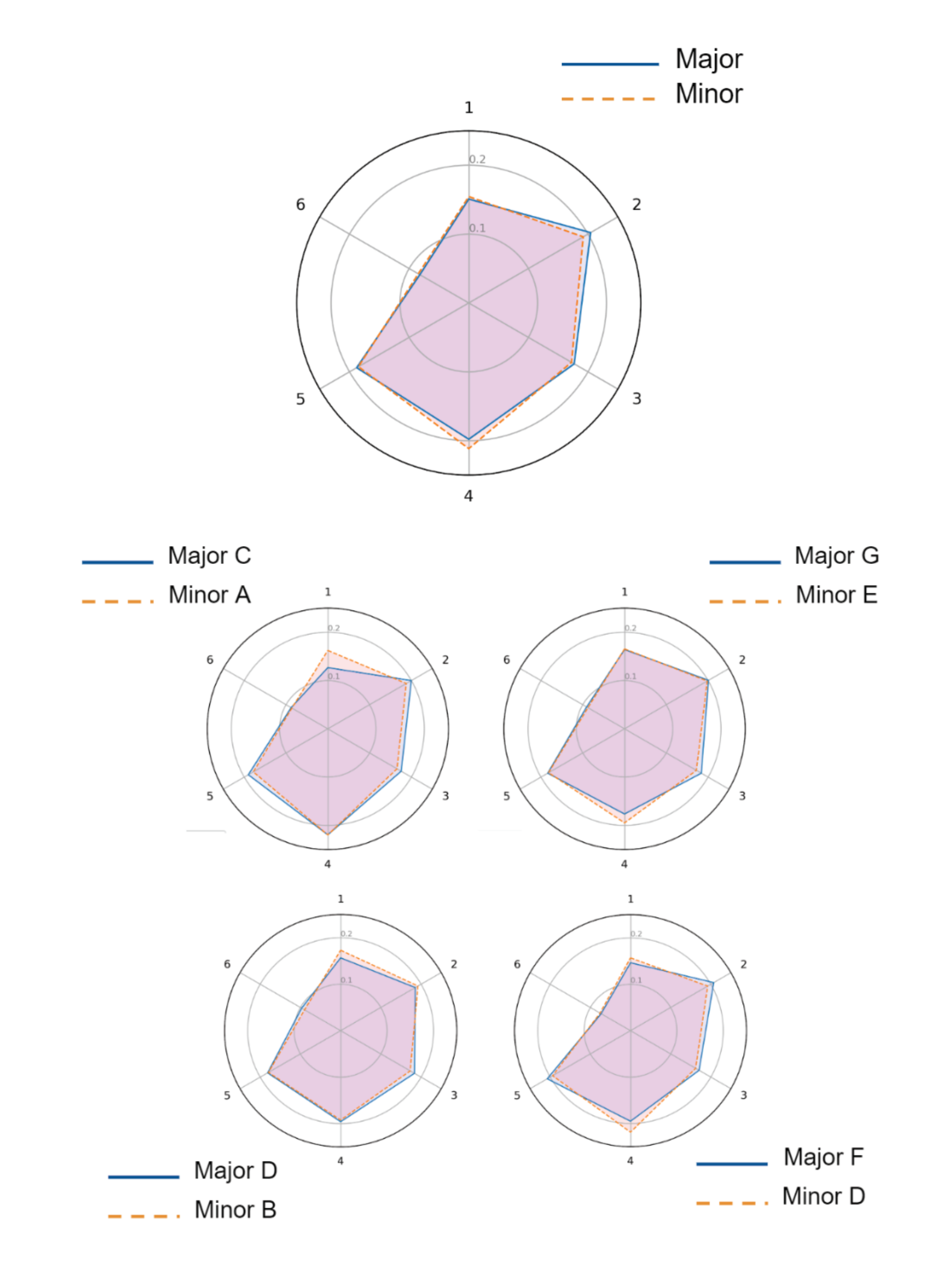}
  \caption{Interval vectors: major and minor}
  \label{fig:radar}
\end{figure}

What is the difference between the interval vector between major and minor? The left panel of Figure \ref{fig:radar} compares interval vectors between the major and minor chords reconstructed by our model. With significance level $0.05$, minor pieces tend to prefer pitches with an interval of four semitones (e.g. $C4$ $E$4), while the major ones with an interval of two semitones (e.g. $C4$ $D4$). For major $C$ and minor $A$, the chord of pitches with an increment of one semitone (e.g. $C4$ $C4\#$) happens more frequently in the minor one, which produces more solemn and ominous sound. It is the same with major scales $G$, $D$, and $F$ compared with their corresponding minor ones $E$, $B$, and $D$. This result also shows that minor scales $A$, $E$, $B$, and $D$ have a weak though significant tendency to use smaller pitch intervals, which is in line with the theoretical review \cite{huron2008comparison}.
\begin{figure*}[t!]
\centering
\includegraphics[width=0.8\linewidth]{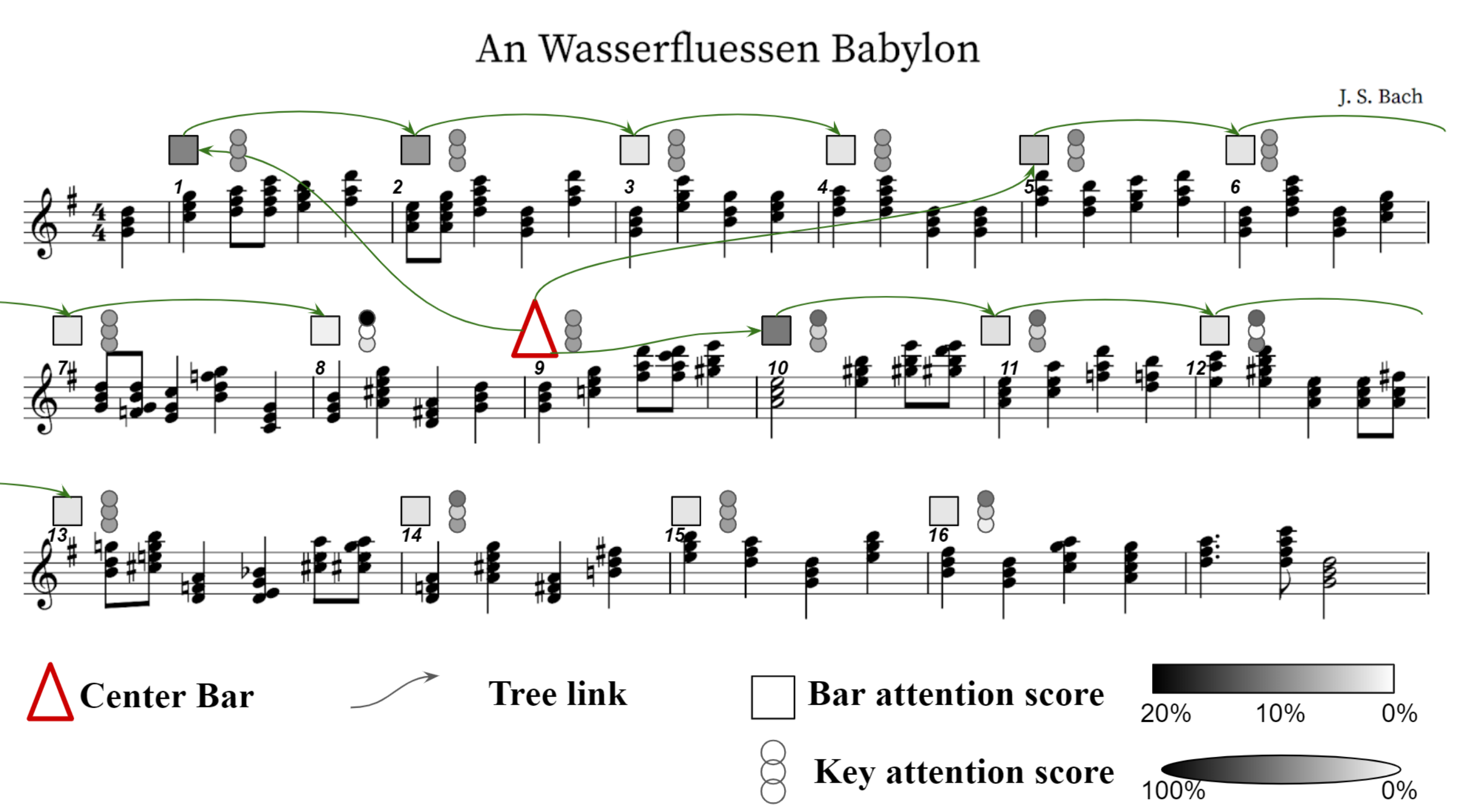}
\caption{The vertical(key) attention and horizontal(bar) attention at the $9$th bar of \textit{An Wasserfluessen Babylon}, with tree link pointing from the parent to its child bar.}
\label{fig:7}
\end{figure*}
\begin{figure}[h]
  \centering
  \includegraphics[width=\linewidth]{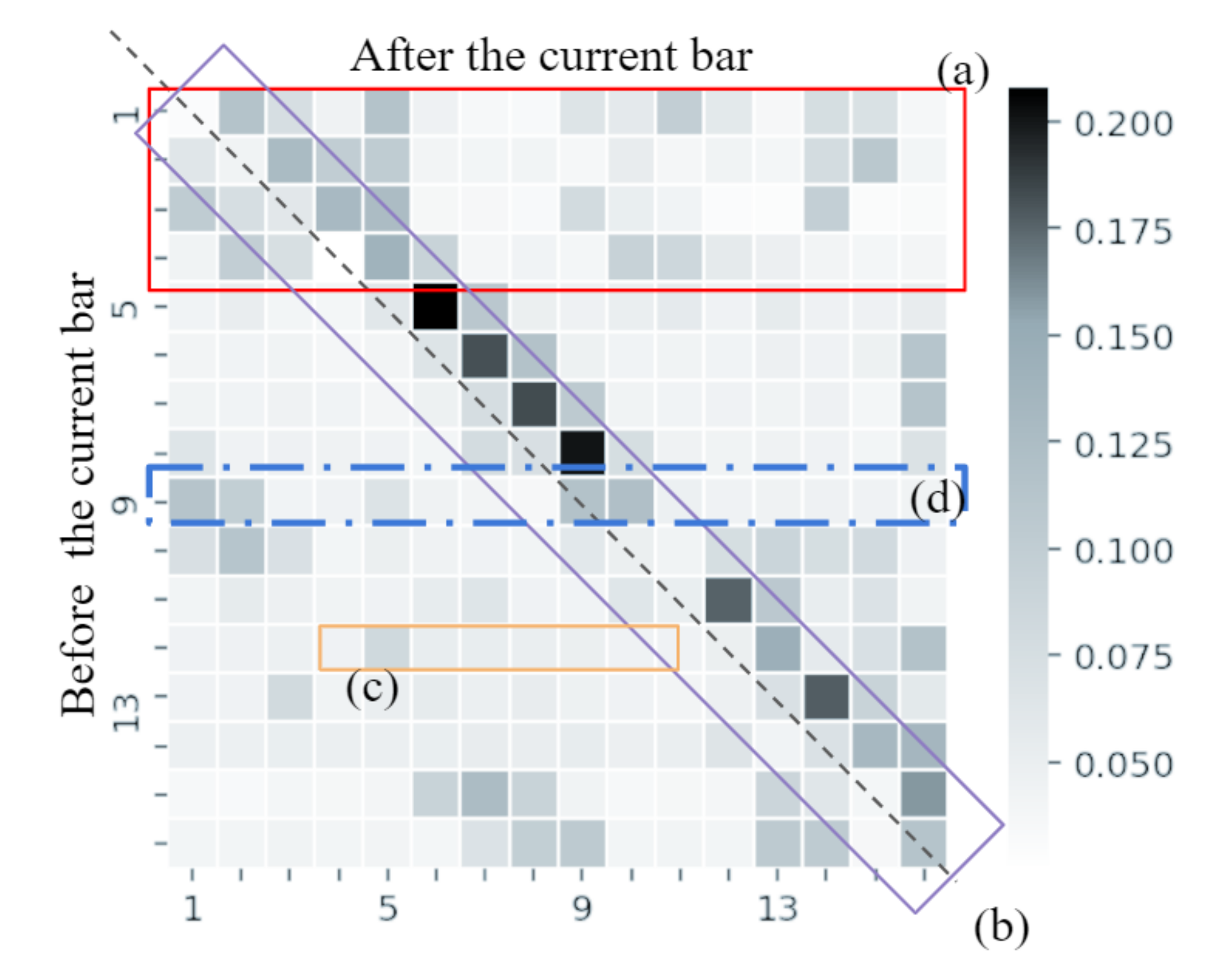}
  \caption{Horizontal attention analysis for \textit{An Wasserfluessen Babylon}: (a) distribution of attention of the beginning bars. (b) coherence afterward v.s. beforehand. (c) rise and fall of attention scores. (d) Structured attention for the $9$th bar (see details in Figure \ref{fig:7})}
  \label{fig:attention_h}
\end{figure}
\subsubsection{Composition}
Music composition is both a craft and an art. One can hardly possess the inspirations and ideas from the composers when enjoying the masterpieces. By using structured attention, our model introduces a novel way to analyze and visualize the structure of musical notes. Notice that in this part we make a modification of our model: the reconstructed pitches not only depend on their previous notes but all notes in the music.
\begin{align*}
        p\left(x_{t} | z_{\leq t}, x_{<t}\right) &\to p\left(x_{t} | z_{\leq T}, x_{<T}\right) \\
        c_{u}^{h}=\sum_{i=1}^{u} P\left(r_{i, u}^{h}=1 | X\right) y_{i} &\to \sum_{i=1}^{N_U} P\left(r_{i, u}^{h}=1 | X\right) y_{i}
\end{align*}
Thus, we can obtain a global tree-structured attention overview for musical pieces. The vertical attention works as a conductor to coordinate keys to generate chords; the horizontal attention provides meaningful suggestions for composition since we can explicitly investigate the relationships between bars. 
Our model teaches several lessons for composition after learning from thousands of piano pieces. First, the beginning few bars have a relatively wide range of attention distribution, which set the tone for the whole music. Second, the current bar usually pays more attention to the consecutive bars after itself than the bars before it, which probably indicates musicians care more about the coherence afterward than the coherence beforehand. Third, the attention has the rise and fall along with rhythms. Typically it increases suddenly to trigger surprise and drops gradually for calming mood. Figure \ref{fig:attention_h} shows a piano piece example from J. S. Bach\footnote{We only consider a total number of 16 bars in the middle by ignoring the beginning anacrusis (pickup measure) and the last one.}. 

Figure \ref{fig:7} depicts the details of the horizontal attention of the $9$th bar and the vertical attention of the first key along bars. Our model shows that the composition of the $9$th pays most attention to the $10$th, $1$st, and $5$th bars, which indicates that it leads the tone of the consecutive bars and matches the meter signature of $\binom{4}{4}$ at bar level. The vertical attention is smooth at most times 
except the $8$th, $12$th, and $16$th bars, suggesting mood or tone changes.

	
    
\section{Conclusion}
This paper proposes a lightweight variant recurrent neural network with vertical-horizontal structured attention to generate music with chords. Our experiment results on MAESTRO dataset show that it can capture not only the temporal relations along time but also the structural relations between keys. We further analyzed the generated music and found our model is sensitive to Western music theory in the sense that it maintains the configuration of the circle of fifths; distinguishes major and minor keys from interval vectors, and manifests meaningful structures between music phases.

\bibliography{main.bib}

\end{document}